\documentclass[runningheads,a4paper]{llncs}

\usepackage{amssymb}
\usepackage{graphicx}
\usepackage{url}
\urldef{\mailsa}\path|talayeh.aledavood@aalto.fi|

\newcommand{\keywords}[1]{\par\addvspace\baselineskip
\noindent\keywordname\enspace\ignorespaces#1}

\begin{document}

\mainmatter  

\title{Channel-Specific Daily Patterns in Mobile Phone Communication}

\titlerunning{Channel-Specific Daily Patterns in Mobile Phone Communication}


\author{Talayeh Aledavood\inst{1}
\and Eduardo L\'opez\inst{2} \and Sam G. B. Roberts\inst{3} \and Felix Reed-Tsochas\inst{2,4} \and\\
Esteban Moro\inst{5} \and Robin I. M. Dunbar\inst{6}\and Jari Saram\"aki\inst{1}}
\authorrunning{Talayeh Aledavood et al.}
\institute{Aalto University School of Science,\\
P.O. Box 12200, FI-00076, Finland\\
\and CABDyN Complexity Center, Sa\"id Business School, University of Oxford,\\
Oxford OX1 1HP, United Kingdom\\
\and Department of Psychology, University of Chester,\\
Chester CH1 4BJ, United Kingdom\\
\and Department of Sociology, University of Oxford,\\
Oxford OX1 3UQ, United Kingdom\\
\and Departamento de Matem\'{a}ticas \& GISC, Universidad Carlos III de Madrid,\\
28911 Legan\'{e}s Spain\\
\and Department of Experimental Psychology, University of Oxford,\\
Oxford OX1 3UD, United Kingdom}

\maketitle

\begin{abstract}

Humans follow circadian rhythms, visible in their activity levels as well as physiological and psychological factors. Such rhythms are also visible in
electronic communication records, where the aggregated activity levels of e.g. mobile telephone calls or Wikipedia edits are known to follow
their own daily patterns. Here, we study the daily communication patterns of 24 individuals over 18 months, and show each individual has a different,
persistent communication pattern. These patterns may differ for calls and text messages, which points towards calls and texts serving a different role in communication. For both calls and texts, evenings play a special role. There are also differences in the daily patterns of males and females both for calls and texts, both in how they communicate with individuals of the same gender vs.~opposite gender, and also in how communication is allocated at social ties of different nature (kin ties vs.~non-kin ties). Taken together, our results show that there is an unexpected richness to the daily communication patterns, from different types of ties being activated at different times of day to different roles of channels and gender differences. 

\keywords{Mobile telephone communication, daily rhythms, computational social science}
\end{abstract}

\section{Introduction}
The human body is equipped with a circadian pacemaker that gives rise to 24-hour rhythms in biological processes within the body, as well as in behavioural patterns~\cite{Kerkhof1985,Czeisler1999,Panda2002}. Studies of human circadian rhythms have traditionally been small-scale studies that involve direct monitoring of human subjects. However, for more than a decade now, automated electronic records of human behaviour have given researchers the ability to study human dynamics and behavioural patterns in unprecedented ways. Circadian rhythms are clearly visible \emph{e.g.} in records of Wikipedia editing~\cite{Yasseri2012}, mobile telephone calls~\cite{Jo2012b,Louail2014} and on Twitter~\cite{Thij2014}. While it is well-known that there is a lot of individual variation in circadian rhythms, these and most other studies of electronic records have focused on aggregate-level phenomena. 
In~\cite{Aledavood2015}, the authors of the present work studied the daily mobile telephone call patterns of individuals and the persistence of such patterns. Here, we expand on this work and also consider another communication channel: text messages.

In~\cite{Aledavood2015}, we showed that individuals have their own distinct daily call patterns, and that these patterns are persistent for each individual even when their social networks undergo turnover. Further, these patterns were seen to have a social dimension: calls at late hours were often associated with close relationships. Because text messages may serve a different purpose in maintaining social relationships than calls (see, \emph{e.g.},~\cite{Lehmann2014,Ling2004}), we address the question of whether the daily patterns of text messaging are similar to those of calls, and whether individuals 
have their distinct, persistent text messaging patterns. Also, because significant differences were seen in call patterns to the same vs.~the opposite gender, as well as kin vs.~friendship ties, we study the daily text messaging patterns from this point of view.

We use the same longitudinal data set of time-stamped text communication records of 24 individuals as in Ref.~\cite{Aledavood2015} (for details, 
see Section~\ref{section:data}). Our results are summarised as follows: first, each individual's text messaging frequency is seen to exhibit distinct daily patterns that are persistent over time, similarly to calls. However, the text messaging and call patterns may differ significantly for a given individual, and on average, text messages are sent more frequently at later hours of the day. Since there is a high level of social network turnover in the studied data set, the persistence of daily patterns for both communication channels indicates that these patterns are not explained in terms of preferred communication timings or channels with specific alters, but rather they contain a component intrinsic to each individual. Further, the difference between the channels is exemplified by daily entropy patterns: even though both calls and texts are targeted at a less diverse set of alters at the late hours, the clear correlation between calls to closest alters and the least diverse times of day is missing for text messages. 

Regarding gender differences, we observed that the total number of text messages is about 1.5 times higher for females than males (for calls, the numbers are practically the same). At the same time, both genders have similar daily trends, sending out the largest numbers of texts in the evening. Calls to kin and family are overall much less frequent than to calls to friends and acquaintances, and text messages to kin are even more infrequent. However, females communicate with their kin by text around 3 times more frequently that males do.

\section{Data}\label{section:data}
For this study, we have used a longitudinal data set of 18 months of auto-recorded, time-stamped phone calls and text communication records of 24 individuals (``egos'' in the following). This data set has been used earlier in ~\cite{Roberts2011,Saramaki2014,Aledavood2015}. Altogether, this data set consists of 74,124 calls and 273,501 text messages, with time stamps at a resolution of one second. The large number of texts compared to calls may have to do with the  young age of participants 
($\sim$18 years at the beginning of the study), as well as with conversations via text messages generating a large number of messages, whereas a conversation via a phone call leaves one record only. As the original purpose of collecting this data was to study turnover in social networks of individuals, the setting was chosen such that all  participants were in their last year of high school at the beginning of the study, and later went either to work or university (often in another city), after about 6 months of data collection. The participants also took part in 3 surveys, separated by 9 months, designed to provide complementary information about members of their communication network (``alters'' in the following). Information on gender and kinship as well as data on how emotionally close egos felt to their alters were collected with these surveys. For further details, please see~\cite{Roberts2011}.

\section{Results}

\begin{figure*}[!ht]
\begin{center}
\includegraphics[width=1.0\linewidth]{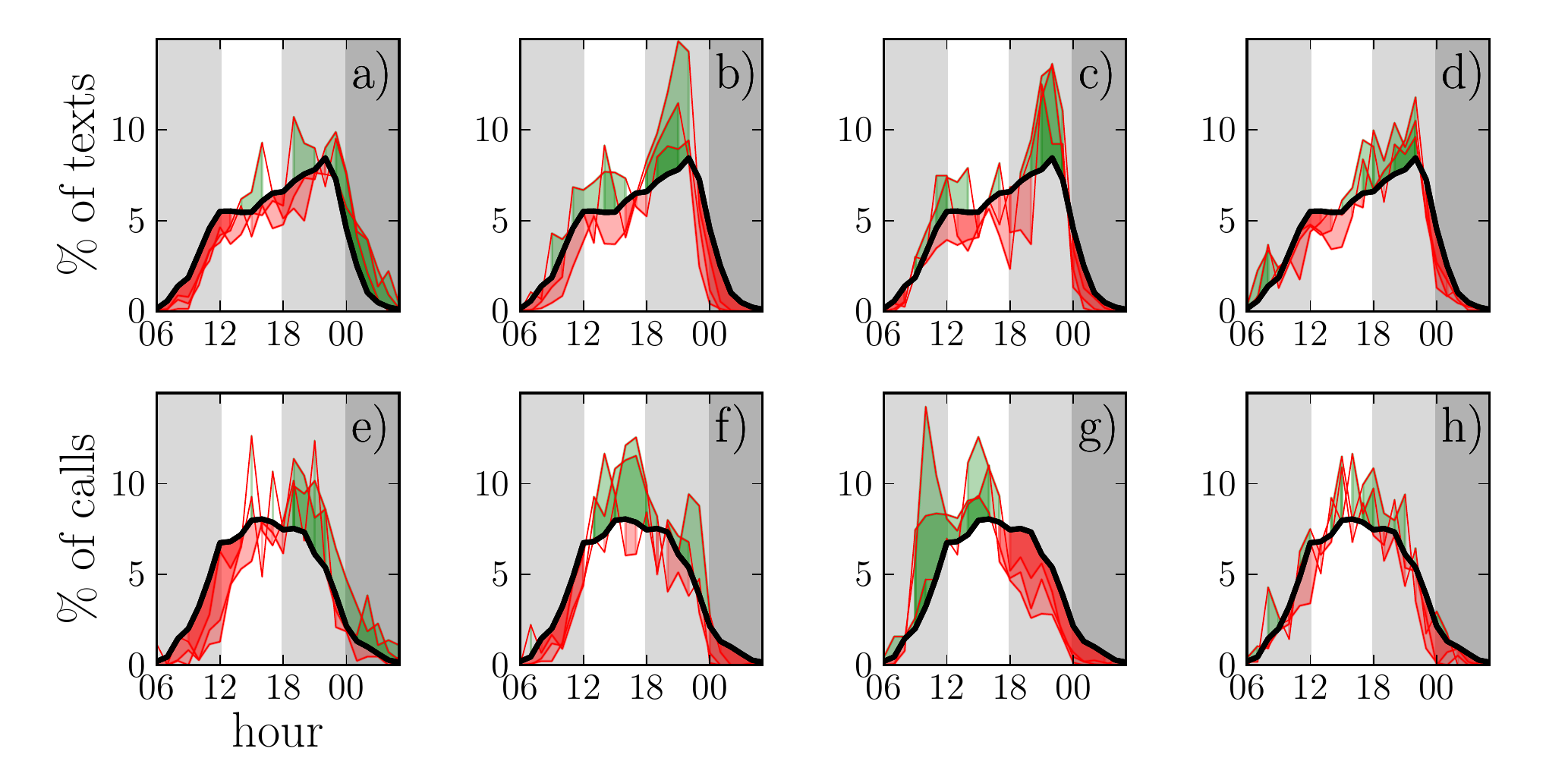}
\caption{(a-d) The daily text patterns of 4 individuals, (e-h) the daily call patterns of the same individuals. The average fraction of calls/texts at each hour of day is denoted by red lines. These have been computed for each of the three 6-month intervals, $I_1$, $I_2$, and $I_3$. The average call and text patterns, averaged over the patterns of all 24 individuals, are shown as black lines. Green shading indicates where an individual's call/text fraction is above average, whereas red shading indicates the opposite. Note the persistence of individual patterns (overlap of green/red areas for an individual) as well as 
the differences between call and text patterns.}
\label{fig:daily_patterns}
\end{center}
\end{figure*}

\subsection{Channel-specific daily patterns and their persistence}

In order to compute the daily patterns of texting for each individual, we begin by segmenting our data temporally. We make two temporal divisions: first, 
at the level of days, we divide each day to 24 one-hour bins and compute the number of calls and text messages inside each bin. In order to address the persistence of the observed patterns, we divide the 18-month time span of our data into three six-month intervals, $I_1$, $I_2$, and $I_3$. The end of the first time interval $I_1$ coincides with early autumn, where the participants move on in their lives and begin work or studies at university. The second time interval, $I_2$, then spans a time range where a major change has taken place in the participants' lives and they are settling in a new environment with major turnover in their social networks.

For each ego, we aggregate all events (calls or texts) within each time interval ($I_1$, $I_2$ or $I_3$) to the hourly bins. To arrive at the daily text or call patterns measuring frequency as function of time, we then sum up and normalise the numbers of respective events at each hour of day. This is repeated separately for each 6-month interval, and each ego. Daily text and call patterns calculated in this manner are displayed in Fig.~\ref{fig:daily_patterns} for four different individuals, together with averages over all 24 individuals. It can be seen that each individual has their own distinct text and call pattern, and both patterns appear fairly persistent over time. It is also evident that the call and text patterns may significantly differ for a given ego. Likewise, there is a clear difference between the average daily patterns of calls and texts: the frequency of text messaging peaks at later hours of the day, whereas the majority of calls are made in the afternoon. This supports the notion of calls and text messages serving different social and communication functions.  

\begin{figure*}[!ht]
\begin{center}
\includegraphics[width=0.75\linewidth]{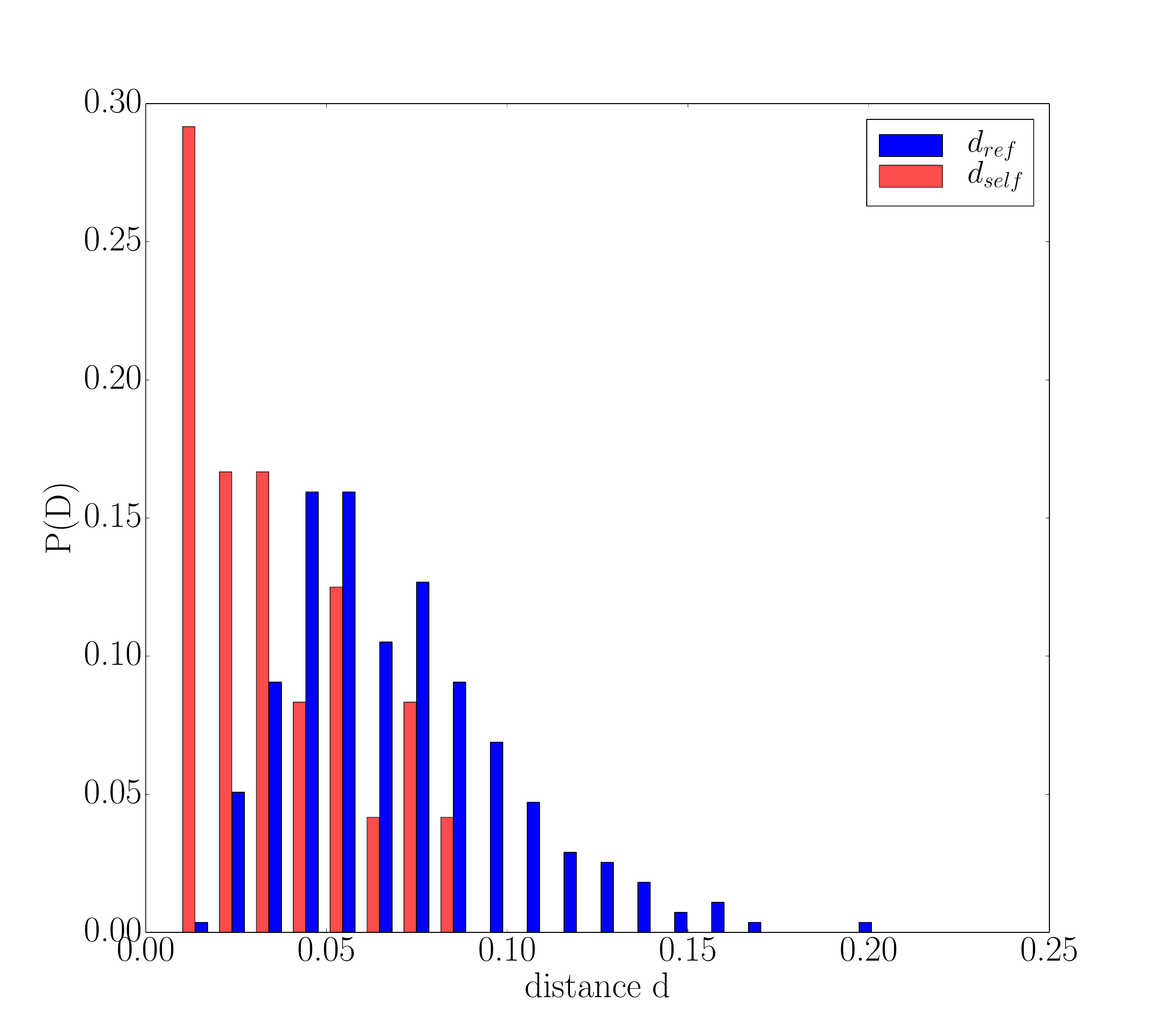}
\caption{Distributions of the values of the Jensen-Shannon divergence, measured between each individual ego's daily text patterns in different 6-month intervals 
($d_\mathrm{self}$) and between patterns of different egos ($d_\mathrm{ref}$). Self-distances $d_\mathrm{self}$ are mostly lower than the reference distances, indicating that each individual's daily text patterns preserve their shape through the 6-month intervals. }

\label{fig:JSD}
\end{center}
\end{figure*}

For quantifying the persistence of each individual's daily text patterns, we apply the Jensen-Shannon Divergence (JSD), similarly to previous works~\cite{Saramaki2014,Aledavood2015}. The JSD is a measure of the dissimilarity of two probability distributions. It is an extension of the Kullback-Leibler divergence (KLD), with the important difference that it can be used for discrete probability distributions with zero-valued elements. For two discrete probability distributions $P_1$ and $P_2$, the JSD is defined as
\begin{equation}
\mathrm{JSD}(P_1,P_2) = H\left(\frac{1}{2}P_1+\frac{1}{2}P_2\right)-\frac{1}{2}[H(P_1)-H(P_2)], 
\end{equation}
where $H$ is the Shannon entropy, $H(P)=-\sum p(t)\log p(t)$. Here, we set $P_i=\{p_i(t)\}$, where $t$ indicates the (binned) time of day, and $i=1,2$ denotes the two distributions to be compared (\emph{e.g.} the two distributions corresponding to $I_1$ and $I_2$ for one ego).
We calculate the self-distance $d_\mathrm{self}$ for each ego as the average of the distances between daily patterns for intervals $I_1$ and $I_2$ and for  intervals $I_2$ and $I_3$. For a reference distance $d_\mathrm{ref}$ with which to compare these values, we calculate the distance between the daily patterns of each ego and all other egos (within the same time intervals),  repeating this for all pairs of individuals and all time intervals. The result can be seen in 
Fig.~\ref{fig:JSD}. It is evident that on average the self-distances $d_\mathrm{self}$ are smaller than the reference distances $d_\mathrm{ref}$, indicating that each individual's daily patterns are fairly persistent. The same was observed for calls in~\cite{Aledavood2015}, but here the differences between self and reference distances are even more evident, \emph{i.e.} daily text patterns appear to retain their shape even better than call patterns.
\subsection{Specifity in communication: who is contacted and when?}

Studies of call records with the data set at hand have revealed a social dimension within the daily patterns: for calls, the diversity of called individuals is on average lower in the evenings and especially at night~\cite{Aledavood2015}. When the called alters are ranked on the basis of the number of calls, it is seen that the fraction of calls to top-ranked alters is often high when the diversity is low; typically, evenings and nights are ``reserved'' for top-ranked alters. 

Here, we set out to study whether similar effects can be detected with text messages (note that as seen in Fig.~\ref{fig:daily_patterns}, calls and texts may follow different daily cycles). We approach the problem using relative entropies as in~\cite{Aledavood2015}: first, we measure the diversity of called/texted alters in the 6-hour bins (6AM-12AM (morning), 12AM-6PM (afternoon), 6PM-12PM (evening) and 12PM-6AM (night)), by computing bin-wise call/text entropies for each ego and interval. These are then normalised by the average entropies computed with a null model, where all called alters are randomly shuffled among calls for one ego (see Methods for details). This null model corresponds to the hypothesis that given the cumulative numbers of calls/texts to each alter and the overall daily pattern, there are no preferred times of calling/texting.

\begin{figure*}[!t]
\begin{center}
\includegraphics[width=\linewidth]{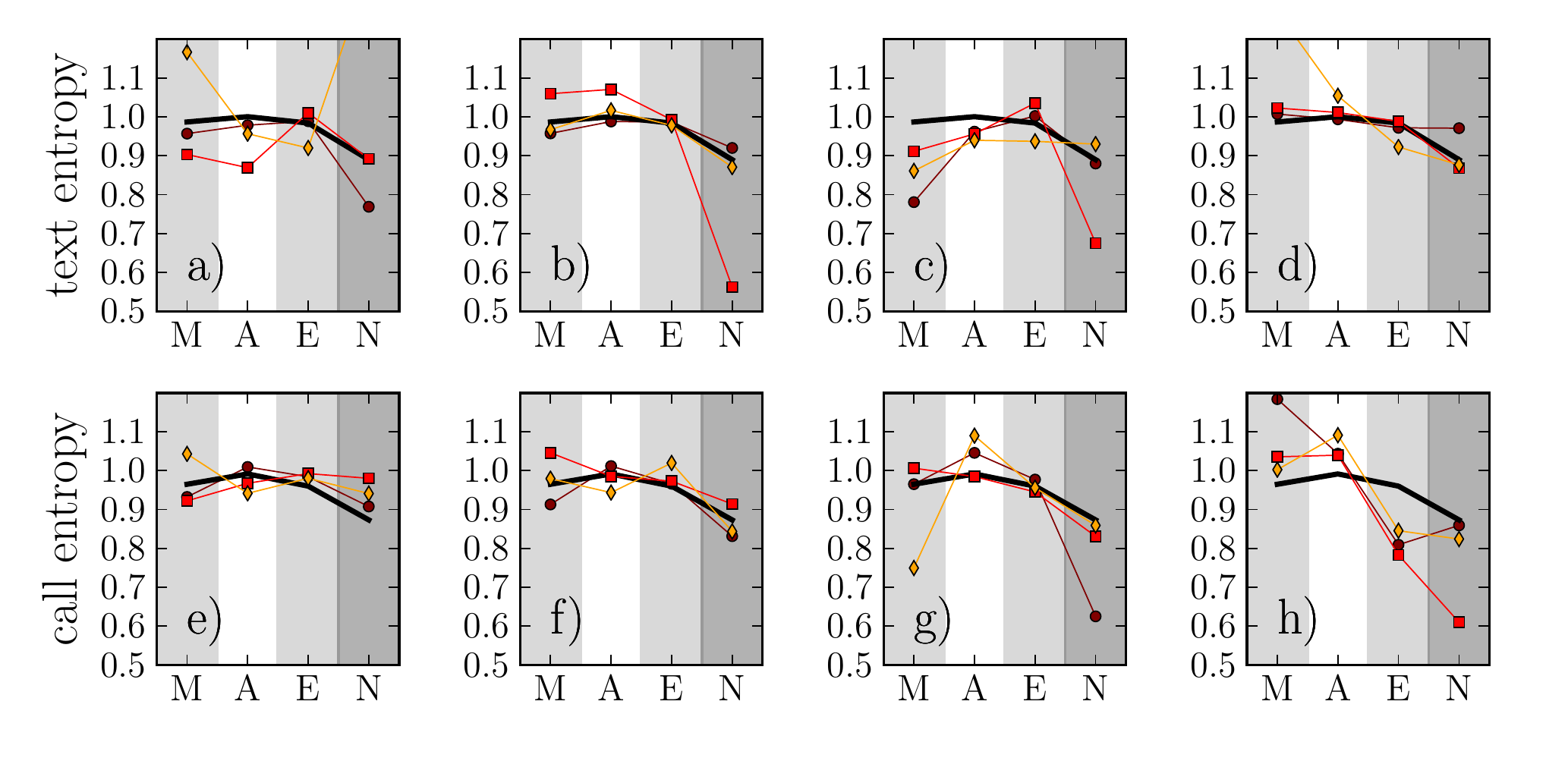}
\caption{The relative entropies for the same 4 individuals as in Fig.~\ref{fig:daily_patterns}. (a-d) Relative entropies for text messages, (e-h) relative entropies for calls. The black line denotes average over all 24 individuals; the coloured lines correspond to the different 6-month intervals for each individual. Time periods are 6AM-12AM (M,orning), 12AM-6PM (A,fternoon), 6PM-12PM (E,vening) and 12PM-6AM (N,ight).}
\label{fig:entropies}
\end{center}
\end{figure*}

As seen in Fig.~\ref{fig:entropies}, the average relative entropies for texting follow a similar pattern as the call entropies, the only difference being that the pattern is slightly more flat. Thus, similarly to calls, text messages in the afternoon are targeted at a more diverse subset of alters -- the relative entropy close to unity indicates that there are no specific preferences. To the contrary, at night and to a smaller extent in the evening, text messages are frequently sent to a specific subset of alters. Note that there is a lot of variation in the individual entropy patterns.

In ~\cite{Aledavood2015}, it was seen that low entropy is often associated with calls to top-ranked alters. We computed the correlation coefficients between relative entropy and fraction of texts to top 3 alters separately for each ego (to avoid the ecological fallacy problem). Unlike for calls, only 7 out of 24 correlation coefficients had a $p$-value less than 0.05, and out of those, 6 coefficients displayed negative correlations. Hence, unlike for calls, communication focused at top-ranked alters do not necessarily explain the low-entropy time ranges. This may have to do with text messages serving a different role in communication, as also seen in the daily frequency patterns. However, 6 out of 7 statistically significant correlation coefficients were still clearly negative, averaging at $r\approx-0.75$, so for certain individuals, a high level of communication to top-ranked alters at certain hours explains the entropy variation.

\subsection{Gender differences in communication patterns}

Next, we turn to gender differences in the daily communication patterns. Overall, when comparing the total numbers of calls and texts, we observe that there is a considerably higher total number of texts than calls, both for males and females. This may have to do with the study participants being about 18 years of age, as heavy users of text messaging are often found in the younger age groups. Further, carrying out a single conversation via text messages may involve a large number of texts. 

\begin{figure}

\centering
\includegraphics[width=0.9\linewidth]{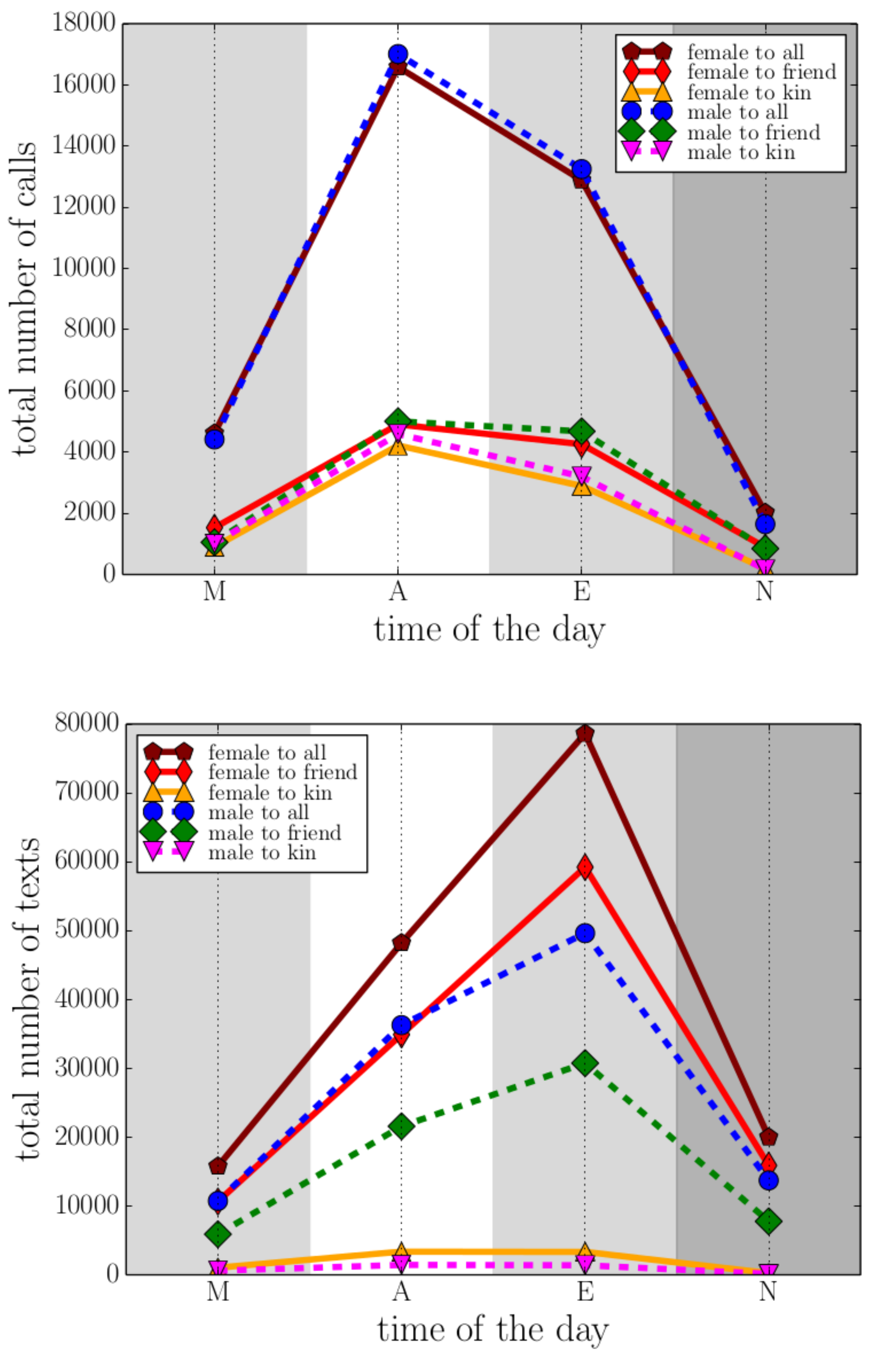}
\caption{Distribution of the total number of calls and texts (top and bottom panels, respectively) at different times of day, for calls/texts by female and male egos to alters of different types. The ``Female to all'' and ``Male to all'' categories contain those unknown alters who have not been recalled in surveys, and for whom no personal data is available}.

\label{fig:FM_total_calls_texts}

\end{figure}

We also see that calls and texts have a different average daily pattern, with calls peaking in the afternoon and texts in the evening (see Fig.~\ref{fig:FM_total_calls_texts}). Frequent texts in the evening may be related to youth culture and communication conventions, and may also have to do with the intrusiveness of the channel; as seen in~\cite{Aledavood2015}, calls at late hours are often targeted at a small subset of closest alters. 
Comparing the communication patterns of males and females, we see that the total numbers of calls made by males and females during the 18 months of data collection are almost equal, whereas females tend to text much more frequently than males. This difference is largest in the evening (see Fig.~\ref{fig:FM_total_calls_texts}). Overall, the total number of text messages is about 1.5 times higher for females than males.

Focusing on different types of social ties, we see that even though the number of calls to friends and kin are similar, for texts they are very different\footnote{Note that in reality the total numbers of calls to friends might be much higher, because for the majority of alters it is unknown whether they are friends, acquaintances, or social ties of some different type. Here, we call those alters for whom an emotional closeness score is available in the surveys ``friends''. However, we can still compare the ratio of calls to friends vs. kin with texts to friends vs. kin, because the set identified as friends is the same both for texts and calls.}. This shows that calling is the dominant channel for communicating with kin. Despite the low numbers of texts to kin both by male and female egos, females in this study have texted their kin about 3 times more often than males, which agrees with other studies that males and females indeed make use of mobile telephones differently~\cite{Zainudeen2010,DeBaillon2005,palchykov2012}.

\section{Summary and conclusions}

We have studied patterns of communication via mobile telephone calls and text messages, and have shown that like many other types of human activity, these patterns follow daily rhythms. Interestingly, the daily patterns of texts and calls appear different and persistent for each individual. Furthermore, the two patterns may significantly differ for a given individual, pointing out that calls and texts may serve different functions. 

One way of interpreting the observed patterns is that they are a superposition of common and unique patterns. First, humans naturally follow the day-night cycle, which is reflected in communication frequency. Second, on top, there may be social conventions and age-group-related effects that individuals typically follow: \emph{e.g.} it is OK to text someone late in the evening, but calls can be made only to one's closest alters. Third, we have individual differences in personality, communication habits and social habits that give rise to each individual's distinct pattern: note that because of the high level of social network turnover in our data, these cannot be explained by communication conventions with specific alters. 

Returning to the differences between calls and texts, our results point out that when studying social networks, data comprising communication along one channel only does not provide a full picture of the network, especially when using the temporal networks framework~\cite{HolmeSaramaki2012}: different channels play different roles, at different times of day. 

\section{Acknowledgements}

TA and JS acknowledge support from the Academy of Finland, project ``Temporal networks of human interactions''  (no. 260427), and computational resources by Aalto Science IT. RD's research is supported by an ERC Advanced grant. SGBR and RD acknowledge support from the UK EPSRC and ESRC research councils for collecting the data.
\bibliography{ECCS_proceeding}

\end{document}